\documentstyle[preprint,aps]{revtex}
\begin{document}
\title{ The dimer-hole-RVB state of the 2-leg t-J ladder: 
A Recurrent Variational Ansatz}

\author{Germ\'an Sierra\thanks{
On leave from  
Instituto de Matem\'aticas y  F\'{\i}sica Fundamental,
C.S.I.C., 28006 Madrid, Spain. em: sierra@sisifo.imaff.csic.es}}
\address{Institute for Theoretical Physics, University of
California, Santa Barbara, CA 93106
}
\author{Miguel Angel 
Mart\'{\i}n-Delgado\thanks{em: mardel@eucmax.sim.ucm.es}}
\address{ Departamento de F\'{\i}sica Te\'orica I, Universidad
Complutense, 28040 Madrid, Spain
}
\author{Jorge Dukelsky\thanks{em: emduke@iem.csic.es}}
\address{Instituto de Estructura de la Materia, C.S.I.C., 
28006 Madrid, Spain
}
\author{Steven R. White\thanks{em: srwhite@uci.edu}}
\address{Department of Physics and Astronomy, University of California, Irvine, CA
92697
}
\author{D.J.\ Scalapino\thanks{em: djs@spock.physics.ucsb.edu}}
\address{
Department of Physics,
University of California,
Santa Barbara, CA 93106
}
\date{\today}
\maketitle

\begin{abstract}
We present a variational treatment of the ground state of
the 2-leg $t$-$J$ ladder,
which combines the dimer and the hard-core boson models into one effective
model. This model allows us to study the local structure of the hole pairs
as a function of doping.
A second order recursion relation is used 
to generate the variational wave function, which substantially 
simplifies the computations.
We obtain good agreement with numerical density matrix
renormalization group results for the ground state
energy in the strong coupling regime. 
We find that the local structure of the pairs
depends upon whether the ladder is slightly or strongly dopped.
\end{abstract}


\section*{Introduction}

The 2-leg, $t-J$ ladder represents one of the simplest 
systems which exhibits some of the 
phenomena associated with high $T_c$ cuprate
superconductivity \cite{DRS,BDRS,GRS,WNS,TTR94,TTR96}.  
The ground state of the undoped system, a 2-leg
Heisenberg ladder, is a spin liquid with a finite spin gap and
exponentially decaying antiferromagnetic spin-spin correlations. Upon
doping, the spin gap remains and there
appear power law CDW and singlet superconducting pairing correlations.
In addition, the pairing correlations have an internal
$d_{x^2-y^2}$-like symmetry with a relative sign difference between the
leg and rung singlets which make up a pair.  Despite all of the
numerical and analytical work which has been done on this system, we
still lack a picture of the ground state which accommodates all of these
physical properties.  There are, however, many hints of what that
picture may look like.  It is the purpose of this paper to take one
step further in that direction.

Short-range resonating valence bonds (RVB) provide a useful basis for
representing the ground state of spin liquids\cite{KRS,LDA}.
For the $t-J$ ladder, 
a $0^{\rm th}$-order picture 
has been provided by the study of the
strong coupling limit where the exchange coupling constant
along the rungs, $J'$, 
is much larger than any other scale in the problem.
The other coupling constants of the model are, $J$: the exchange
coupling constant along the legs, and $t$ and $t'$: the hopping
parameters along the legs and the rungs respectively.  
In the limit $J' >> J, t, t'$, 
the ground state  of the undoped ladder is simply given
by the coherent superposition of singlets across the rungs.
Addition of one hole requires the breaking of one of these singlets,
in which case the hole gets  effectively 
bound to the unpaired spin, becoming a quasiparticle
with spin $1/2$ and charge $|{\rm e}|$. Addition of another hole 
leads to the binding of two holes in the same
rung in order to minimize the cost in energy. 
In this picture there is no
spin-charge separation, a fact that remains valid down to intermediate
and weak couplings, as confirmed by various numerical and analytical
studies. 
Based on this  picture 
it is possible to construct an effective theory
describing the motion and interactions
of the hole pairs \cite{TTR96}. It is given by  a 
hard-core boson model (HCB)  
characterized by an effective hopping parameter $t^*$ and
interaction $V^*$ of the hole pairs.
The HCB model
describes  the 
doped ladder as a Luther-Emery liquid,
with gapped spin excitations and gapless charge collective modes,
which are responsible for the CDW and SC power law correlations. 
We summarize the $0^{\rm th}$-order picture in Figure 1, which 
shows a typical state of HCB's, as well as the two
building blocks that are used its construction. 

In order to go beyond this picture, we need  to consider the fluctuations
of the states of the HCB model. 
To lowest order in perturbation
theory they are shown in Fig.~2.
The admixture of the 
state shown in Fig.~2(a)  is of order $J/J'$ and  represents a  resonance 
of two nearest neighbor rung singlets.
According to the standard RVB scenario,
this resonance
effect  leads to a substantial
lowering of the ground state energy. 
The state in Fig.~2(b) is of order $t/J'$, and it  
can be though of as a bound state
of two quasiparticles, whose  characteristic feature is the diagonal
frustrating bond across the holes. From the RVB point of view,
Fig.~2(b) is a resonance of a singlet and  a hole pair. 
The importance of this state, even for intermediate
couplings such as $J=J'=0.5 t$, was emphasized in
the DMRG study of reference \cite{WS}, where it was shown to be the most 
probable configuration of two dynamical holes in a 2-leg ladder. 
In the HCB model of \cite{TTR96},
the states of the form
of Fig.~2(b) are taken into account
as intermediate or virtual
states, which lead to the effective hoping, $t^*$ 
and interaction, $V^*$ between the hole pairs.
It is clear however that 
``integrating out"  the diagonal states through perturbation 
theory, erases the internal structure of the hole pairs. Here we want to
extend the HCB description to include the internal structure 
of the hole pairs.

In order to define an effective model which would
retain the degrees of freedom associated with  the internal structure
of the hole pairs, we need to consider the states that appear in second
order in the strong coupling expansion.
They are given 
in Fig.~3. Let us comment on them. The state of Fig.~3(a) is of order
$(J/J')^2$ and it is a  higher order RVB state,
whose  contribution to the ground state of the undoped ladder was studied
in \cite{SMD}. In this reference it was shown that its inclusion
in a variational ansatz improves the numerical results,
but does not change the qualitative picture obtained using 
the  dimer ansatz \cite{WNS,FM}.
The state of Fig.~3(b), which is in fact first order in $t'$,
can be seen as a bound state of two quasiparticles, 
while 3(c) and 3(d) are higher order
corrections to the diagonal state shown in Fig.~2(b). 
For these reasons it seems 
consistent to keep the state 3(b) 
on an  equal footing with the states 2(a) and  
2(b). To give further
support to  this choice, we notice that the exact solution for two
holes on the $2\times 2$ cluster requires a superposition of the states
shown in 2(b) and 3(b) along with 1(a) and 1(b)
(see Fig.~4) \cite{WS}. 

In summary, we conjecture that in order to discuss the nature
of the superconducting order parameter of the doped 2-leg, $t-J$ ladder, 
in the strong coupling regime, it is sufficient to consider
states built up from 5 possible local configurations,
given by  
rung-singlet-bonds (Fig.~1(a)), rung-hole-pairs (Fig.~1(b)),  
two-leg-bonds (Fig.~2(a)), hole-pairs with a singlet diagonal
bond (Fig.~2(b)) and  hole-pairs with a singlet leg bond
(Fig.~3(b)). A typical state
constructed using  these building blocks is shown in Fig.~5. 
We shall call these types of states {\it dimer-hole-RVB}  states.
The effective model that governs their dynamics will be called
the {\it dimer hard-core boson model}  (DHCB) and its Hamiltonian
can  be determined by considering the fluctuations
of the dimer-hole states, in a manner similar to the one considered
above for the HCB states. The DHCB model contains
spin and charge degrees of freedom, 
together with their couplings, and in that sense is an interesting
model to study the interplay between the two types of degrees
of freedom, although here we will
focus on the variational ground state of the model. 

The mathematical formulation of the
DHCB model involves
an interesting but complicated
combination of vertex and Interaction Round a Face (IRF)
models. The latter terminology is borrowed from 
Statistical Mechanics \cite{B}.
The vertex variables
describe  the number of electrons per rung, i.e. $n_i =0,1,2$,
while the IRF variables describe the number and type 
of bonds connecting
two rungs, i.e. $\ell_{i,i+1} = 0, 1_d, 1_h, 2$, where  the subindices
$d$, $h$ indicate the diagonal or horizontal nature of the bond.
The only allowed configurations
for two consecutive IRF 
variables  $(\ell_{i,i+1}, \ell_{i+1,i+2})$ 
are: $(0,0), (1_d,0), (1_h,0), (2,0)$ together with their permutations.
Moreover the vertex variables are subject to certain constraints
imposed by the IRF ones. Namely, A) if $\ell_{i,i+1}= 1_d $ or $1_h$
then $n_i = n_{i+1} = 1,$ and  B) if $\ell_{i,i+1}= 2$ 
then $n_i = n_{i+1} = 2$. Only if $\ell_{i,i+1}=0$ can $n_i$ 
and $n_{i+1}$ take any value, i.e. 0, 1 or 2. 

It is beyond the scope of this work  to  present a full account 
of the DHCB model. Instead, we shall try to uncover some 
of its physics, by means of a combination of two
approaches, namely the Density Matrix Renormalization
Group\cite{W} and the Recurrence Relation Method (RRM)\cite{SMD}. 
While the DMRG is a powerful numerical technique,
which in many cases yields the 
exact answer, the RRM is essentially analytic, 
lacking  the numerical precision of the DMRG, but
sharing with it some features, as for example the
Wilsonian way of growing the system by the addition of sites
at the boundary. In the RRM one begins with an assumption 
about the local configurations through which the system grows.
Then  one may test whether the state that is generated 
gives results in agreement 
with the essentially exact DMRG results.

\section*{ The variational wave function}

The Hamiltonian of the 2-leg, $t-J$ ladder is given by,

\begin{eqnarray}
& {\cal H}  = {\cal H}_S + {\cal H}_K = 
\sum_{\langle i,j \rangle} J_{ij} \;
( {\bf S}_i \cdot {\bf S}_j - \frac{1}{4} n_i n_j) & \nonumber \\
&-   \sum_{\langle i,j \rangle , s} t_{ij} \; P_G \; 
( c^\dagger_{i , s} c_{j,s} 
+ c^\dagger_{j , s} c_{i,s} ) \; P_G &
\label{1}
\end{eqnarray}

\noindent
where $J_{ij}, t_{ij} = J, t$
or $J', t'$, depending on whether the link $\langle ij \rangle$ is along
the legs or the rungs respectively.  $P_G$
is the Gutzwiller projection operator which forbids double occupancy.
The rest of the operators appearing in (\ref{1}) are standard (we use the
conventions of reference \cite{WS}). Each site $i$ is labelled by the 
coordinates $(x,y)$ with  $x= 1, \dots , N$ and $y= 1,2$. We choose
open boundary conditions along the legs of the ladder.

The pair field operator which creates a pair of electrons, at the sites
$i$ and $j$, out of the vacuum is given by,

\begin{equation}
\Delta^\dagger_{i,j} = \frac{1}{\sqrt{2}} 
( c^\dagger_{i , \uparrow} c^\dagger_{j,\downarrow} 
+ c^\dagger_{j , \uparrow} c^\dagger_{i,\downarrow} ) 
\label{2}
\end{equation}

As explained in the introduction, we want to built up
an ansatz for the ground state based on the 5
local configurations of the DHCB model. The explicit realization
of these configurations  in terms of 
pair field operators are given by ( see Fig.6),

\begin{equation}
\begin{array}{rl} 
| \phi_{1,1} \rangle_x = &  |0 \rangle_x \\
| \phi_{1,0} \rangle_x = & \Delta^\dagger_{(x,1) (x,2)}\,\, |0\rangle_x \\
| \phi_{2,0} \rangle_{x,x+1} =& -u\,\, \Delta^\dagger_{(x,1) (x+1,1)} \,\,
\Delta^\dagger_{(x,2) ( x+1, 2)} \,\, |0\rangle_{x,x+1}  \\
| \phi_{2,1} \rangle_{x,x+1} = & 
[ b\,\, ( \Delta^\dagger_{(x,1) (x+1, 2)}
+  \Delta^\dagger_{(x,2) ( x+1, 1)} ) \\
& + c\,\, 
( \Delta^\dagger_{(x,1) ( x+1, 1)} +  \Delta^\dagger_{(x,2) ( x+1, 2)} ) ]
\,\, |0 \rangle_{x,x+1}
\end{array}
\label{3}
\end{equation}

\noindent where $|0\rangle_x$ is the Fock vacuum associated with the
rung labelled by the coordinate $x$
( $|0\rangle_{x,x+1} = |0\rangle_x \otimes |0\rangle_{x+1} $).
The states $|\phi_{n,p} \rangle$,
involve $n=1,2$ rungs and $p=0,1$ pairs of holes. 
The variational parameter $u$ gives the amplitude of the
resonance of a pair of bonds between vertical and horizontal positions 
\cite{SMD}, while $b$ and $c$ are the variational parameters
associated with the diagonal and horizontal configurations of two
holes respectively. In the strong coupling limit,  
$J' >> J, t, t'$, 
we expect to find $ u \sim J/J', \,\,b \sim t/J' $ and 
$ c \sim t t'/J'^2$.

Let us call $|N,P \rangle$ the ground state of a ladder with $N$ rungs and
$P$ pairs of holes. Of course we should be in a regime of the 
coupling constants where there is binding of two holes. 
The state  $|N,P \rangle$ will  be in general a linear superposition
of the dimer-hole states of  Fig.5, which suggests  that
working with this sort of states could be a formidable task. 
Fortunately, we can apply  
the method developed in \cite{SMD} to generate $|N,P \rangle$ 
in a recursive manner, in terms of 
the states of the ladders with $N-1$ and 
$N-2$ rungs, and $P$ and $P-1$ pairs of holes.   
In \cite{SMD} it was shown that 
$|N ,P =0 \rangle$, which is in fact 
a dimer-RVB state \cite{WNS,FM}, can be
generated by a second order recursion relation. Then by a simple procedure  
one can compute overlaps and expectation values of different
operators using recursion formulas, whose thermodynamic limit
can be studied analytically.

Following the strategy of considering first the HCB states
and then the DHCB ones, we shall give the rule that generates the former
type of states. It is given by the first order recursion relation,

\begin{equation}
|N+1,P+1 \rangle = |N,P+1 \rangle \, |\phi_{1,0} \rangle_{N+1}
+ \,  |N,P \rangle \, |\phi_{1,1} \rangle_{N+1}
\label{4}
\end{equation}

\noindent supplemented  with the initial conditions,

\begin{eqnarray}
& |1,0 \rangle = |\phi_{1,0}\rangle & \nonumber \\ 
& |1,1 \rangle = |\phi_{1,1}\rangle & \label{5} \\
& |N, P \rangle = 0,   \,\,\,{\rm for} \,\, N < P & \nonumber
\end{eqnarray}

Calling $F^{HCB}_{N,P}$ the number of linearly independent states
contained in $|N,P\rangle$, we deduce from  Eq.(\ref{4}) the recursion
relation,

\begin{equation}
F^{HCB}_{N+1, P+1} = F^{HCB}_{N, P+1} + F^{HCB}_{N, P}
\label{6}
\end{equation}

\noindent whose solution is given by the combinatorial number,

\begin{equation}
F^{HCB}_{N,P} = \left( \begin{array}{c} N \\ P \end{array} \right)
\label{7}
\end{equation}

\noindent Eq. (\ref{7}) is the dimension of the Hilbert space of 
the  HCB model with $N$ sites and $P$ pair of holes.
We have not introduced variational parameters in Eqs. (\ref{5}),
but if we did, then  all states of the Hilbert
space of the HCB model would be generated by the first order recursion
relation. 
It may be worthwhile to recall that the HCB model is essentially
equivalent to the spinless fermion model or the XXZ model \cite{TTR96}.

Turning now to the DHCB model, the key point  is 
to realize that
the dimer-hole states can be generated
by the following  second order  recursion relation,
involving the local configurations given by eq.(\ref{3}),

\begin{eqnarray}
& |N+2, P+1 \rangle =  |N+1, P+1 \rangle \,\, | \phi_{1,0} \rangle_{N+2} 
& \nonumber \\
& +   |N+1, P \rangle \,\, | \phi_{1,1} \rangle_{N+2} +
|N, P+1 \rangle \,\, | \phi_{2,0} \rangle_{N+1,N+2} +
|N, P \rangle \,\, | \phi_{2,1} \rangle_{N+1,N+2} & \label{8} 
\end{eqnarray}

\noindent with the  initial conditions (\ref{5}). 
See Fig.7 for a graphical representation of (\ref{8}).

{\bf Counting dimer-hole states}

Let $F_{N,P}$ denote the number of dimer-hole states of a 
2-leg ladder with $N$ rungs containing $P$ pairs of holes.
According to (\ref{8}) they satisfy the  recursion relation

\begin{equation}
F_{N+2,P+1} =F_{N+1,P+1} + F_{N,P+1}+ F_{N+1,P} +4 F_{N,P}
\label{9}
\end{equation}
with the initial conditions

\begin{equation}
F_{N,N} = 1, \,\, F_{N,P} =0 \,\,\, {\rm for } \,\,\, \, N < P
\label{10}
\end{equation}

From (\ref{9}) and (\ref{10}) we deduce that $F_{N,0}$ satisfies the
well known Fibonacci recursion formula \cite{SMD}, 
and that in the limit of very
large $N$ it grows  exponentially,

\begin{equation}
F_{N,0} \sim \Phi_0^N, \,\,\,( N>>1)
\label{11}
\end{equation}

\noindent where $\Phi_0 = \frac{1}{2} ( 1 + \sqrt{5})$ is the
golden ratio. 
Using generating function methods\cite{SMD} 
one can  easily solved the recursion relation 
(\ref{9}), together with the initial condition (\ref{10}). The result
is given by the contour integral,

\begin{equation}
F_{N,P} = \oint  \frac{d z}{2 \pi i} 
\frac{ z^{N+1} \, ( z +4)^P }{ (z^2 - z -1)^{P+1} }
\label{12}
\end{equation}

\noindent
where the contour encircles the singularities of the integrand.
For $P=0$ the integrand has two simple poles at the zeros
of the polynomial $z^2-z-1$, the largest of which is precisely
the golden ratio $\Phi_0$. In this way one gets Eq.(\ref{11}). 
For a finite number of holes the residue formula applied to (\ref{12})
yields, to leading
order in $N$

\begin{equation}
F_{N,P} \sim N^P\,\, \Phi_0^N, \,\,\,\, N >> 1, \,\,
P : {\rm finite}
\label{13}
\end{equation}
 
\noindent
where the proportionality constant depends only on $P$. 
Let us finally consider  the limit 
where both $N$ and $P$ go to infinity, while keeping their ratio fixed,

\begin{equation}
x = \frac{ {\rm Number }\,{\rm of } \,{\rm holes} }{
 {\rm Number }\,{\rm of } \,{\rm sites}}= 
 \frac{P}{N}\  , \quad 0 \leq x  \leq 1\  
\label{14}
\end{equation}

\noindent Here $x$ can be identified with the hole doping factor
of the state $|N,P\rangle$. The saddle point method
applied to (\ref{12}) 
gives the asymptotic behaviour of the number of dimer-hole
states for a finite density of holes,

\begin{equation}
F_{N,P} \sim f(x)^N ,\,\,\,\,\,\,\, f(x) = 
\frac{ \Phi ( \Phi + 4)^x }{ (\Phi^2 - \Phi -1)^x} 
\label{15}
\end{equation}

\noindent
where $\Phi = \Phi(x)$ is the highest root of the following
equation

\begin{equation}
x = \frac{ (\Phi^2 - \Phi - 1)( \Phi + 4)}{ \Phi ( \Phi^2 + 8 \Phi -3)}
\label{16}
\end{equation}

The function $f(x)$ is depicted in Fig. 8.  
Observe that $\Phi(0) = \Phi_0$. The effect of 
a finite density of holes is that of moving a singularity. 
This phenomena also occurs in the  computation of  the energy, 
and other observables.

{\bf Ground State Energy}

The parameters $u,b,c$ are found by the standard minimization 
of the mean value of the 
energy $\langle N,P| H_N | N,P\rangle/\langle N,P|N,P \rangle $,
where $H_N$ denotes the Hamiltonian of the ladder with $N$ rungs.
The usefulness of Eq.(\ref{8}) is that it implies that the
wave function and energy overlaps also satisfy recursion relations.
Let us define the following quantities,

\begin{equation}
\begin{array}{cl}
Z_{N,P} = & \langle N,P | N,P \rangle \\
Y_{N,P} = & _N \langle \phi_{1,0}| \langle N-1, P| N,P \rangle \\
E_{N,P} = & \langle N,P| H_N | N,P \rangle \\
D_{N,P} = & _N \langle \phi_{1,0}| \langle N-1, P| H_N |N,P \rangle \\ 
W_{N,P} = &  \langle N,P| n_N  | N,P \rangle
\end{array}
\label{17}
\end{equation}

\noindent where $n_N$ is the number operator acting on the rung
$N$.
The off-diagonal overlaps arise from the cross terms when
applying (\ref{8}) to the ket and the bras in 
$\langle N+2,P+1|N+2,P+1 \rangle 
$ and $\langle N+2,P+1|H_{N+2}|N+2,P+1 \rangle $.
The recursion relations satisfied by (\ref{17}) are given by,

\begin{equation}
\begin{array}{rl}
Z_{N+2, P+1} = & Z_{N+1, P+1} + u^2 \,\, Z_{N, P+1} + u \,\,
Y_{N+1, P+1} +  \,\, Z_{N+1, P} + 2(b^2 +c^2) \,\, Z_{N, P} \\
Y_{N+2, P+1} = & Z_{N+1, P+1} + u/2 \,\, Y_{N+1, P+1} \\
E_{N+2,P+1} = &  E_{N+1,P+1} - J' \, Z_{N+1,P+1} + u^2 \,
 E_{N,P+1} - (2 J + J'/2) u^2 \,  Z_{N,P+1} +  \, E_{N+1,P} \\
& + 2 ( b^2 + c^2)  E_{N,P}
- (2 J c^2 +4 b t + 8 b c t') \,  Z_{N,P}
+ u D_{N+1,P+1} \\
&-2 u (J +J'/2) Y_{N+1, P+1} - 4 t b Y_{N+1,P}\\
& - \frac{1}{4} J \, W_{N+1, P+1} - \frac{1}{4} J u^2 W_{N, P+1}
- \frac{1}{4} J ( b^2 + c^2) W_{N,P} \\ 
D_{N+2,P+1} = & E_{N+1,P+1} - J' Z_{N+1, P+1}
+u/2 D_{N+1, P+1} - u (J +J'/2) Y_{N+1, P+1} \\
& - 2 t b Z_{N,P} - \frac{1}{4} J\, W_{N+1, P+1} \\ 
W_{N+2, P+1} = & 2 Z_{N+1, P+1} + 2 u^2 \, Z_{N,P+1}
+ 2 ( b^2 + c^2) Z_{N,P} + 2 u Y_{N+1, P+1} 
\end{array}
\label{18} 
\end{equation}

\noindent The initial conditions read,

\begin{equation}
\begin{array}{ll}
Z_{0,0} =1,&  Y_{0,0} = E_{0,0}= D_{0,0}= W_{0,0} =0 \\
X_{N,P} = 0, & {\rm for }\,\,\,\, N < P \,\,\, {\rm and}\,\,
 X = Z, Y, E, D, W \end{array}
\label{19}
\end{equation}

For finite values of $N$ and $P$, and given choices of $u,b,c$,  
one  can iterate numerically the recursion relation (\ref{18}) 
using the initial conditions (\ref{19}) and look for the 
minimum of the ground state energy
$E_{N,P}/Z_{N,P}$. We give below the results obtained using
this variational
method for a $2\times 32$ ladder and compare them with the corresponding 
results obtained with the DMRG.

\section*{The RRM wave function versus the DMRG: numerical results}

As explained in the introduction the DHCB model is 
the appropiate framework to study the strong coupling limit
of the 2-leg ladder, if one wishes to take into account the
local structure of the hole pairs. To check the validity of
this asumption we have studied the cases where the coupling
constants takes the following values, 
$t=t'=1, J =0.5$ and $J'= 0.5, 1,2,3,4$ and 5. In this manner
we go from the intermediate coupling regime, i.e. $J' \sim 1$ 
to the strong coupling regime $J' \gtrsim 3.5$. We are always
working in a non-phase-separated region.

In Figure 9 we show the ground state energy of the 2 x 32 ladder,
for the previous choices
of parameters, computed with the RRM for all dopings 
and the DMRG for $x= 1/8, 1/2$ and 7/8. One sees that the
results obtained with the RRM wave function  agree reasonably well with
those of the DMRG and their
accuracy improves  as  $J'$ increases.  

The kinetic energy of the ladder is shown in Fig.10. 
It has the pattern expected for a collective charge
mode, as described by the HCB and the DHCB models.
The similarity between this figure and Fig.8
have a common origin. 
They both correspond to
holes moving collectively through the spins in a complicated
many body state.

Fig.10 shows the existence of an optimal doping for
which the kinetic energy is a minimum. 
The existence and position of this optimal doping depends
on the values of the coupling constants.

The nature of this many body state is clarified 
by figures 11, 12 and 13 where we show the values
of the variational parameters $u,b$ and $c$ as functions
of the doping $x$ for different coupling constants. 
The parameter $u$  starts from a positive value
corresponding to the undoped ladder \cite{SMD},  and
it decreases upon doping until a critical value $x_c(J/J')$, 
where it vanishes. For higher dopings $u$ becomes negative. 
For the undoped ladder the parameter $u$ can be interpreted
as the square of the RVB amplitude $h_{{\rm RVB}}$ 
for having a 
bond along the legs\cite{SMD}. 
The analogue amplitude for a bond along the rungs has been
implicitly normalized to 1. For low doping, i.e. $x < x_c$,
since $u(x) > 0$, we can  similarly define a doping dependent
amplitude for a leg-bond as

\begin{equation}
u(x) = h_{{\rm RVB}}^2(x) > 0, \;\;\;\;( x < x_c)
\label{20}
\end{equation}

\noindent
In order to fulfill the 
Marshall theorem for the undoped ladder one
requires the RVB amplitude $h_{{\rm RVB}}(0)$ to be positive
\cite{LDA}, which explains why $u(0)$ 
is also positive. Actually for the positivity of
$u(0)$ one just need  $h_{{\rm RVB}}(0)$ to be a real number.
At $x=0$ $h_{{\rm RVB}}(0)$ increases with 
$J/J'$ due to the resonance between rung and 
leg singlets, according to the RVB scenario. 
Upon doping, however, the holes give rise to  destructive interference
which degrades progressively the aforementioned 
resonance mechanism. 
This explains why $u(x)$ and $h_{{\rm RVB}}(x)$ decrease
with $x$. For $x < x_c$ the ground state is dominated
by the resonating 
valence bonds and the RVB picture remains qualitatively correct.

For $x > x_c$ the interference due to the holes has driven $u$ 
negative and it is no longer appropiate to interpret $u(x)$
as the square of $ h_{{\rm RVB}}$. Rather, 
the physical interpretation of the overdoped region
comes from the
solution of the Cooper problem 
in the $t-J$, 2-leg ladder, and its BCS extension. 
It can be shown analytically that two electrons in the latter system
form a bound state only under certain conditions (details
will be given elsewhere).
For $J=0.5, t=t'=1$ one must have $J' > 3.3048$,
( note that the binding 
of two electrons in the $t-J$ chain 
requires $J/2t >1$ \cite{O}). 
The exact solution for 4 or more electrons is 
difficult to construct,
but we expect it to be given essentially by a 
Gutzwiller projected BCS like wave function.
A short range version of the 
latter type of wave function can be generated
from the recursion relation (\ref{8}), with $u$ a negative
parameter, which can be written as

\begin{equation}
u(x) = - h_{{\rm BCS}}^2(x) < 0, \;\;\;\;( x > x_c)
\label{21}
\end{equation}

\noindent where $h_{{\rm BCS}}$  is the BCS amplitude for 
finding two electrons
at distance 1 along the legs. Of course this interpretation of $u$ as
minus the square of a BCS amplitude requires it to be negative. As we
put more electrons into the ladder the value of $h_{{\rm BCS}}$ 
decreases 
and for electron densities larger than $1 - x_c$, we switch into the
RVB regime.

The difference between the underdoped 
and overdoped regimens can be attributed to two different
internal structures of the pairs. In the low doping regime 
$x < x_c $, holes doped into the spin-liquid RVB state
form pairs with an internal $d_{x^2-y^2}$-like structure
relative to the undoped system. However for $x > x_c$ 
one moves into the low density limit characterized 
by electrons doped into an internal $s$-wave like symmetry.
This issue will be discussed
in detail in a separate publication.

Let us now comment on Figs. 12 and 13. Both  are very similar and show
that for $x \sim 1/2$, $b$ and $c$ 
reach their maximum. At $x=1/2$ 
there are as many electrons as holes, and in a certain sense the 
ground state of the ladder is a large scale reproduction of the microscopic
ground state of the 2 x 2 cluster given in Fig. 4. Indeed for 
$J=J'=0.5, t=t'=1$ the ratio $b/a$ of the parameters appearing in
Fig. 4 is given by 1.30, which is very close to the value of $b$ at its
maximum. For $x < 0.7$ and $J'=0.5$ the parameter $b$ is larger than 1
and it is always larger than $c$ for all dopings and couplings. 
This is in agreement with the DMRG results of \cite{WS}, which
show the importance of the diagonal frustrating bonds above the horizontal
or vertical ones for $J/t=J'/t=0.5$.

Finally Fig.14 is a $J/t-n$ diagram which shows the boundary of
phase separation obtained by means 
of the DMRG and the RRM in the case where $J=J', t=t'=1$. Observe
that this is not the strong coupling case we have been discussing so
far, and hence the validity of the RRM  is more questionable. 
In any case, we see an overall agreement between both results 
(see references \cite{TTR96,HP,S} 
for comparisons with other numerical results). 
In the two-leg $t$-$J$ model, phase separation is controlled by
$J$, rather than $J'$, so the strongest coupling we have considered
above, $J'/t = 5$, $J/t=0.5, t'/t=1$, does not phase separate.

\section*{Conclusions}

In this paper we have proposed an extension of the effective
hard-core boson model (HCB) of the 2-leg ladder 
of reference \cite{TTR96}, in order 
to include the local structure of the  hole pairs. 
The extended effective model, called the DHCB model,    
contains both dimer bonds, hard core bosons
and various combinations between bonds and holes, whose  
relevance  have been studied previously 
with DMRG \cite{WS}. 
Generalizing the methods of reference \cite{SMD}
to the case with holes, we study a variational ansatz
for the ground state of the DHCB model, which
depends only on three variational parameters. 
The resulting dimer-hole state is generated by a second
order recursion formula, which also leads to recursion 
formulas for the overlaps necessary to compute 
the  energy of the ansatz.  We give the 
results of the energy  minimization
for the 2 x 32 ladder and compare them with those obtained with the
DMRG method in the strong coupling region. The recursion relations
we have derived for the ground state energy can be solved analytically
in the thermodynamic limit and the minimization can be then done
numerically.  Finally we give a
physical interpretation of the behaviour of the variational
parameters with doping.

\section*{Acknowledgments}

GS would like to thanks the organizers of the ITP program 
``Quantum Field Theory in Low Dimensions: From Condensed Matter
to Particle Physics"  for the warm hospitality. 
MAMD thanks the organizers of the Benasque Center of Physics 1997
for their support and hospitality.  
GS acknowledges support from the 
NSF under Grant No. PHY94-07194 and the Direcci\'on General
de Ense\~{n}anza Superior, MAMD acknowledges support from the  
CICYT under contract AEN93-0776,
JD acknowledges support from the 
DIGICYT under contract 
No. PB95/0123, SRW acknowledges support from the
NSF under Grant No. DMR-9509945, and  DJS acknowledges support from the
NSF under Grant numbers PHY-9407194 and DMR-9527304.

\newpage 
\section*{Figure captions}

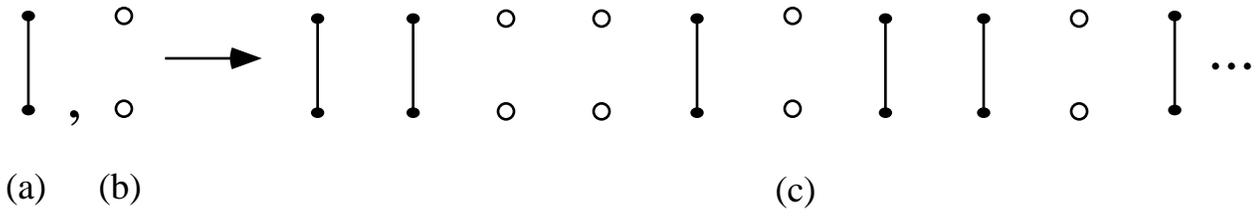
 {\bf Figure 1:}The $0^{\rm th}$  order picture of the Hard Core Boson model:
a) The vertical bond, b) the vertical hole-pair singlet, 
c) a typical state of the HCB model.

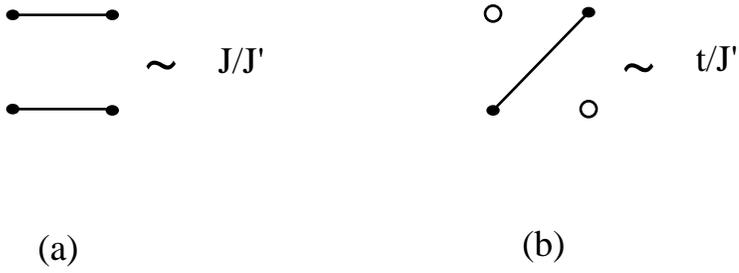
 {\bf Figure 2 :} The two lowest order states in the strong coupling
limit $J^{\prime} \gg J, t, t^{\prime}$ of the HCB model. They represent
the first order contribution to the DHCB model.
a) the resonance of two vertical bonds,
b) bound state of two quasiparticles.

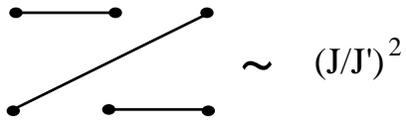
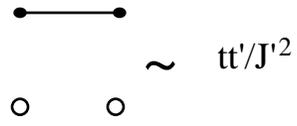
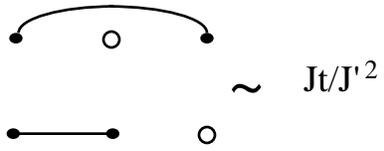
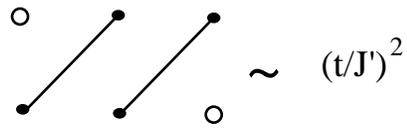
 {\bf Figure 3 :} Higher order strong coupling states contributing to
the DHCB model.
a) a higher order RVB state,
b) a bound state of two quasiparticles,
c) and d) higher order corrections to the diagonal state 2(b)).

 {\bf Figures 4 :} The exact ground state for a single plaquette with
two holes\cite{WS}(case $N=2$ and $P=1$).

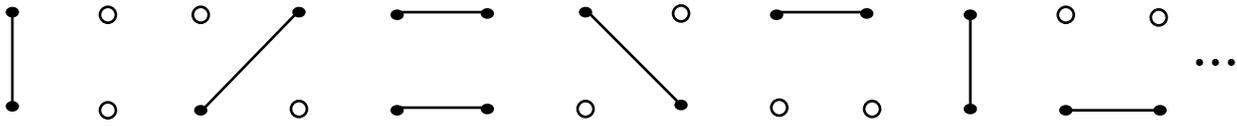
 {\bf Figure 5 :} A typical dimer-hole-RVB state.

 {\bf Figure 6 :}Elementary building block states of the RRM used
in the construction of the dimer-hole states.

 {\bf Figure 7 :}A pictorical representation of Eq. (\ref{8}).

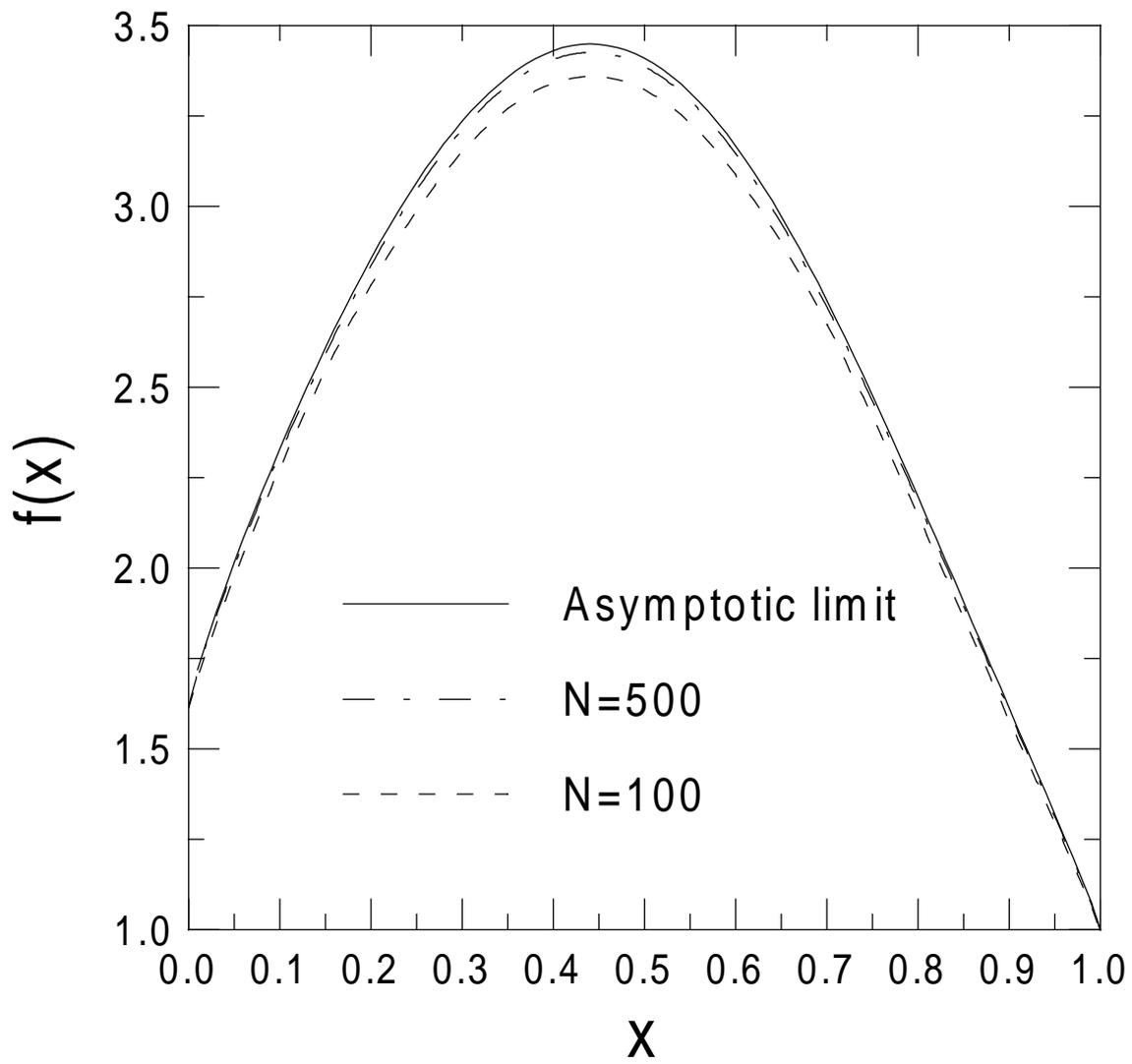
 {\bf Figure 8 :}The function $f(x)$ appearing in (\ref{15}). 
The maximum appears at $x=0.44$.

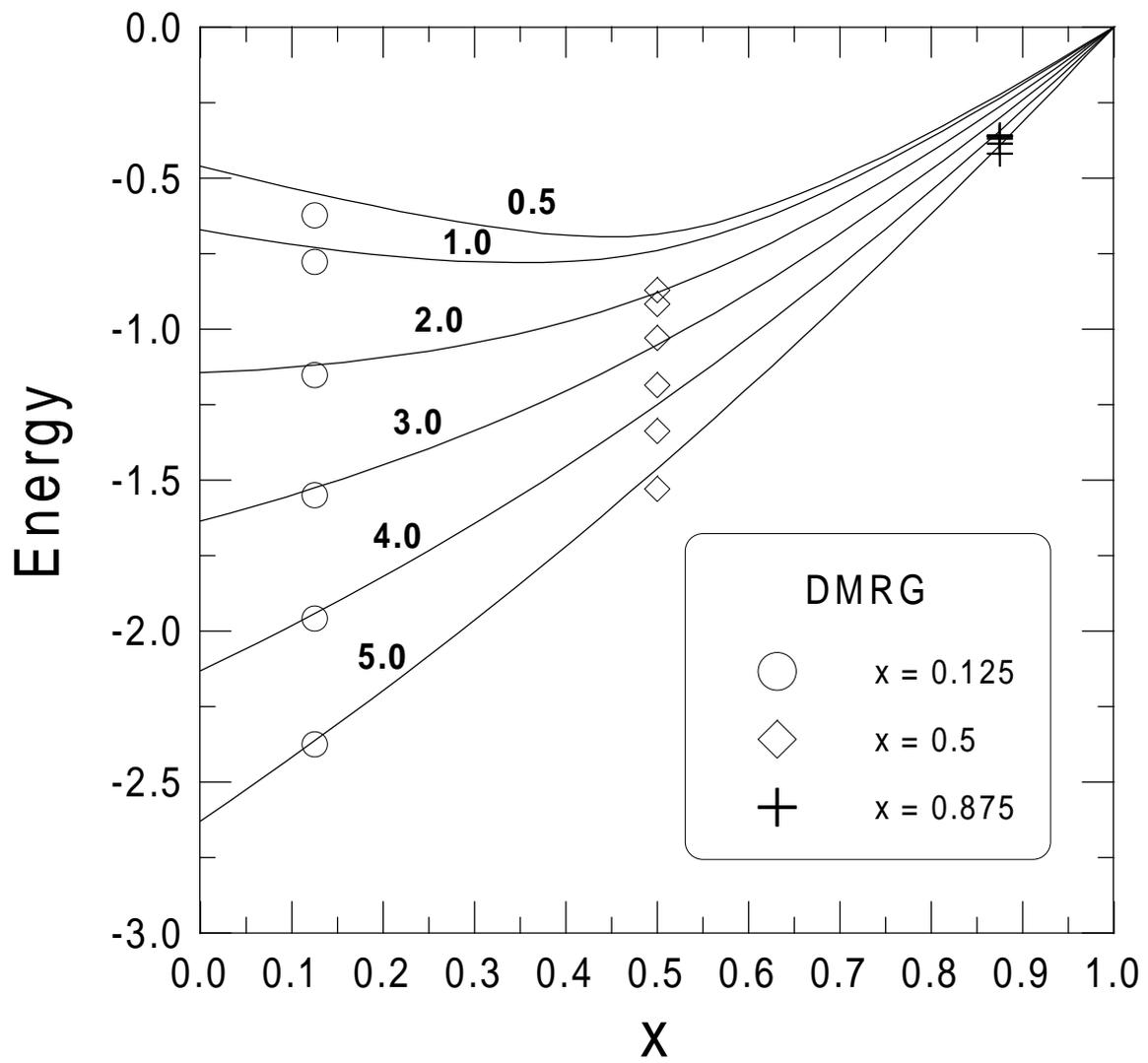
 {\bf Figure 9 :} Ground state energy per site of the 2 x 32 ladder
with $J=0.5, t=t'=1$ and $J'=0.5,1,2,3,4,5$.  The remaining data
given below in figures 10-13 also corresponds to these 
choices of couplings. 
The continuum curves are obtained with the RRM, while the 
special symbols are the DMRG data 
corresponding to $x=1/8, 1/2 $ and 7/8 respectively.

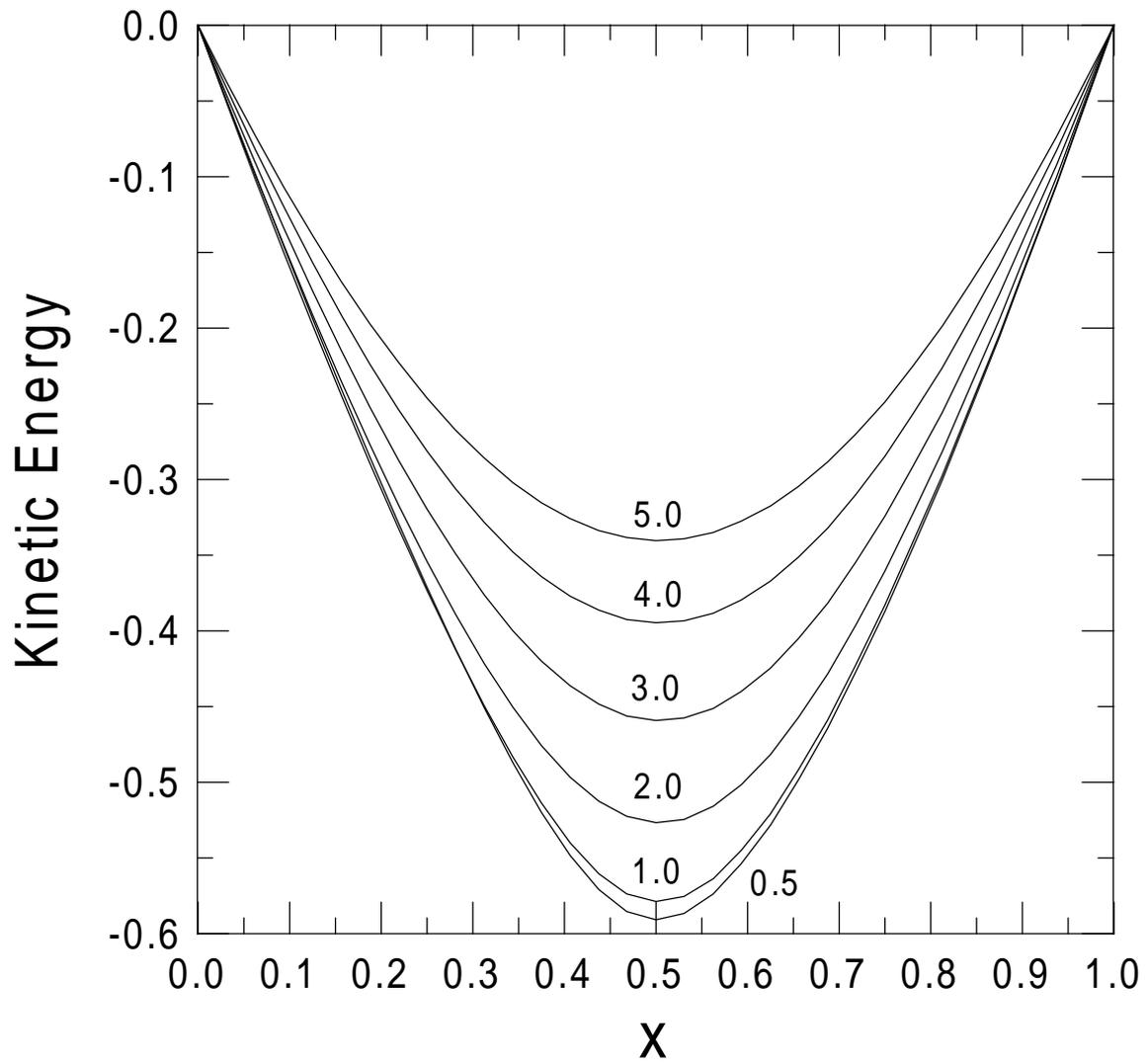
 {\bf Figure 10 :} Kinetic energy per site.

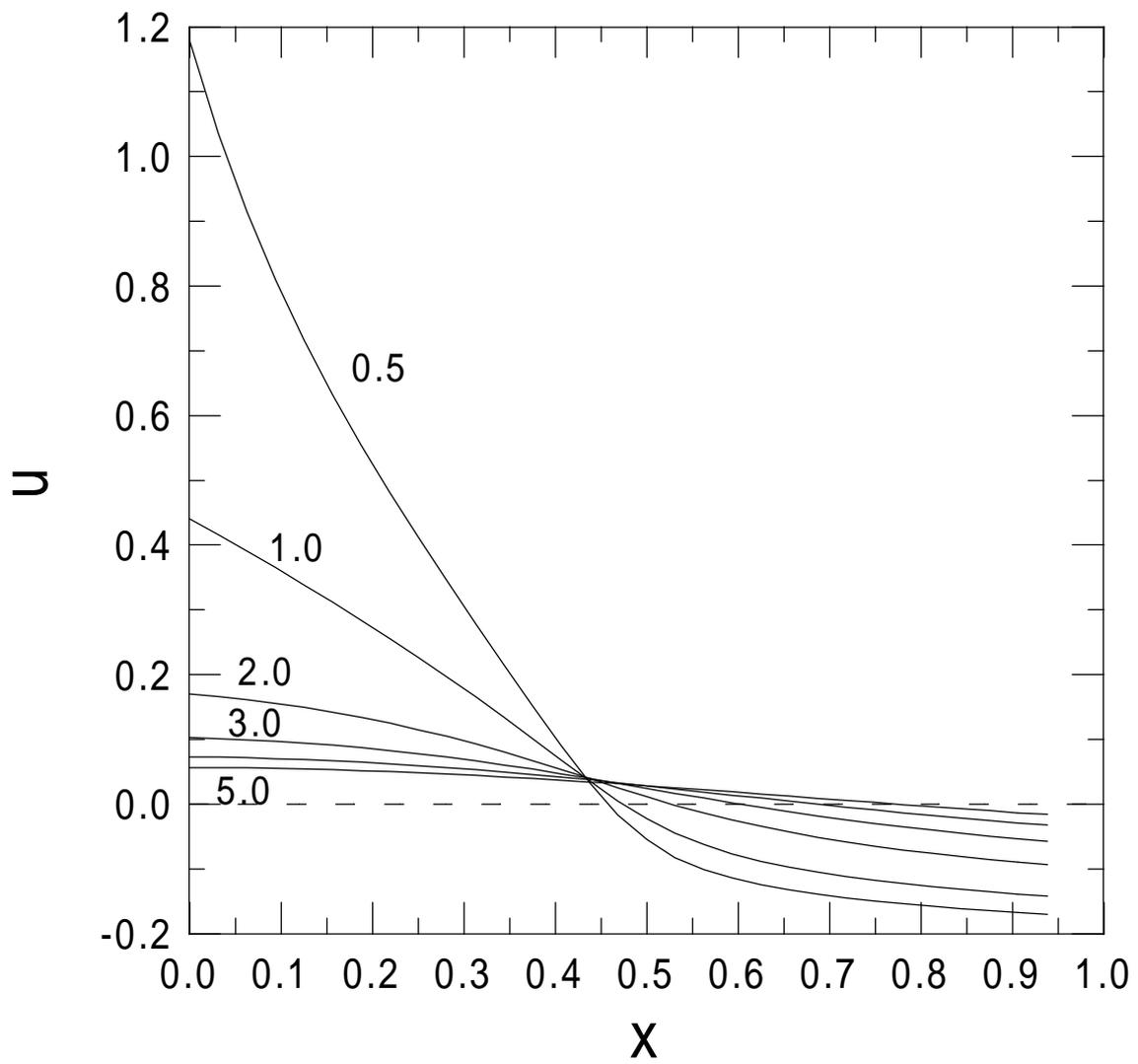
 {\bf Figure 11 :} The variational parameter $u$ as function of the doping.

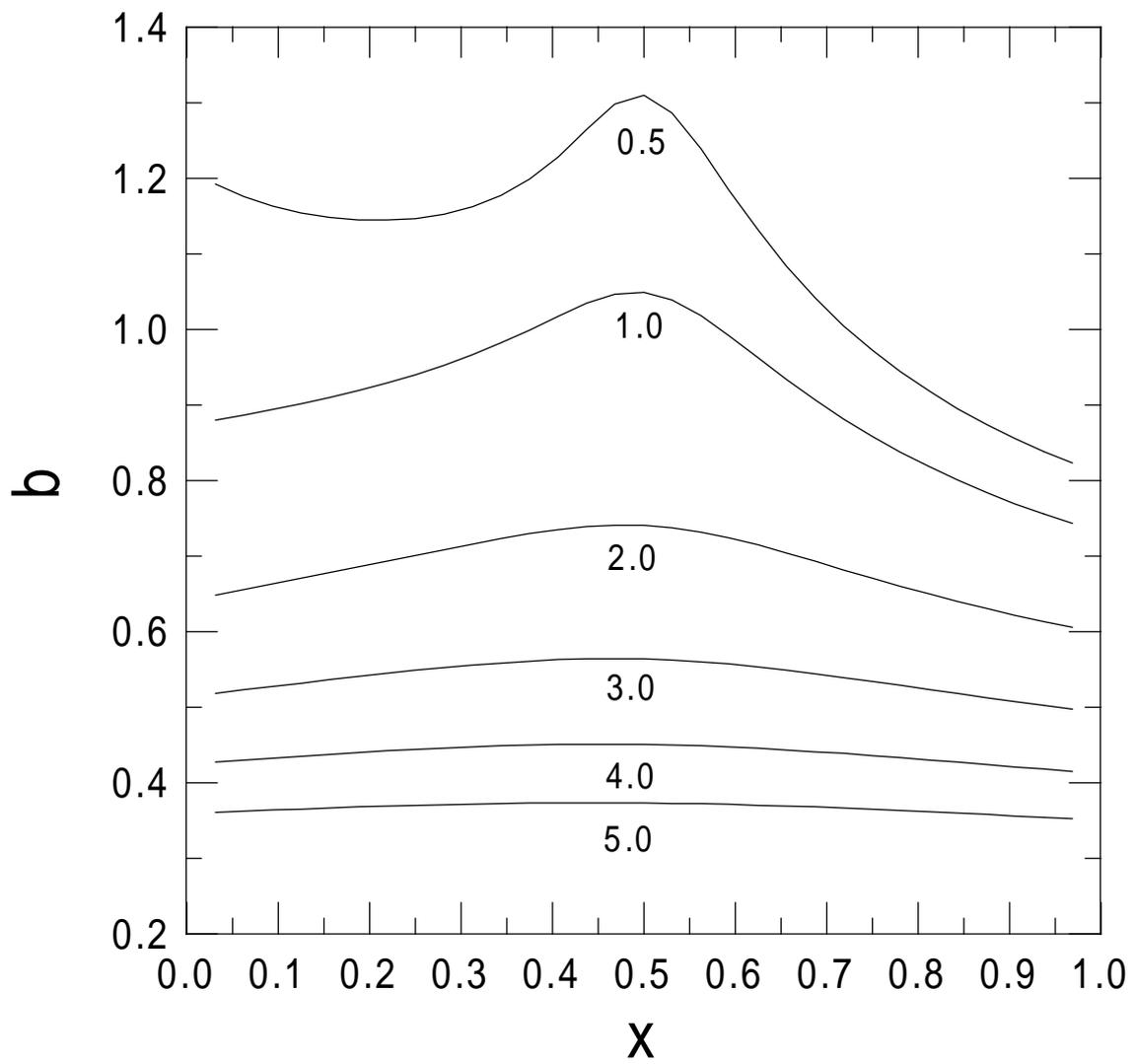
 {\bf Figure 12 :} The variational parameter $b$ as function of the doping.

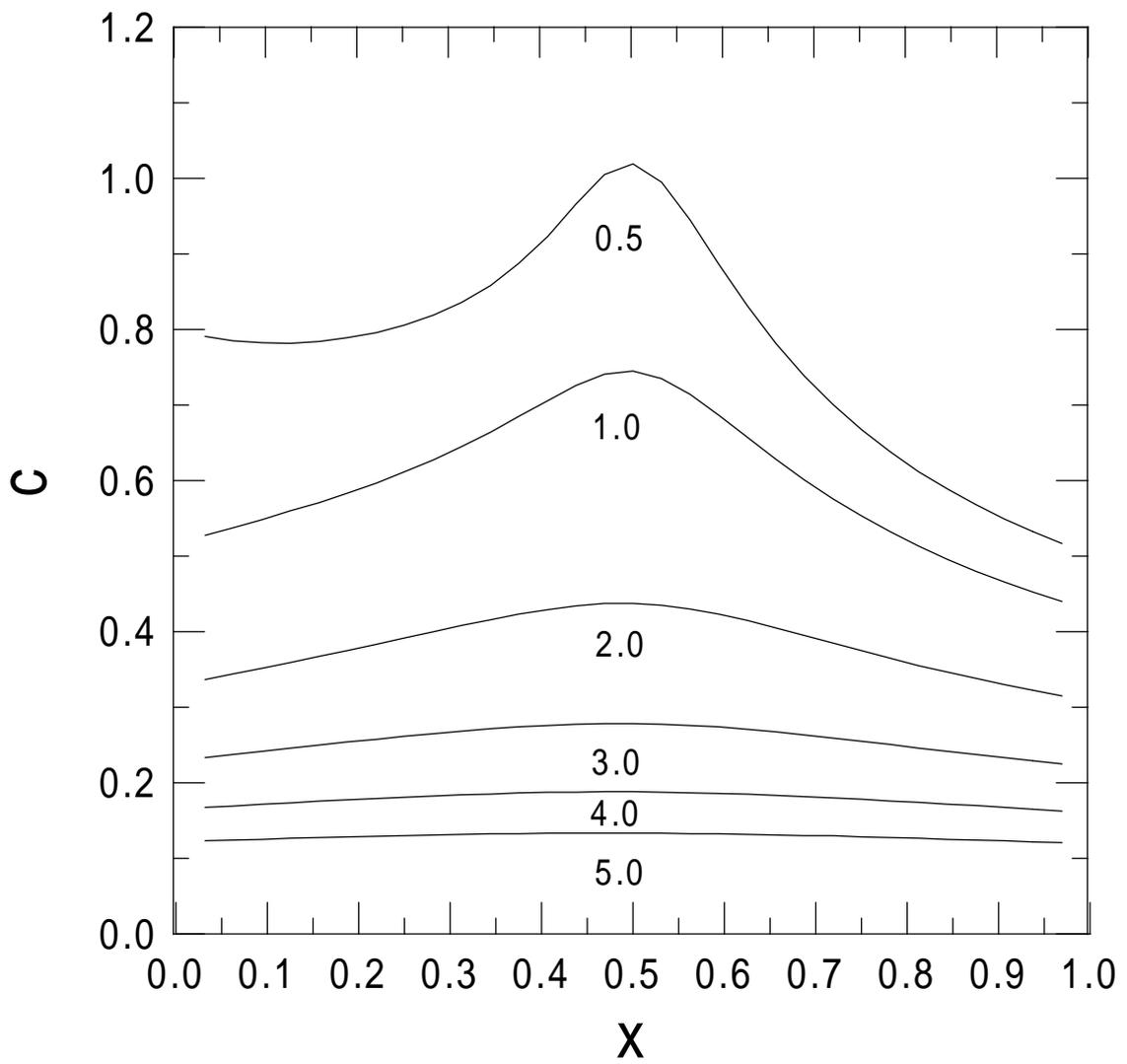
 {\bf Figure  13 :} The variational parameter $c$ as function of the doping.

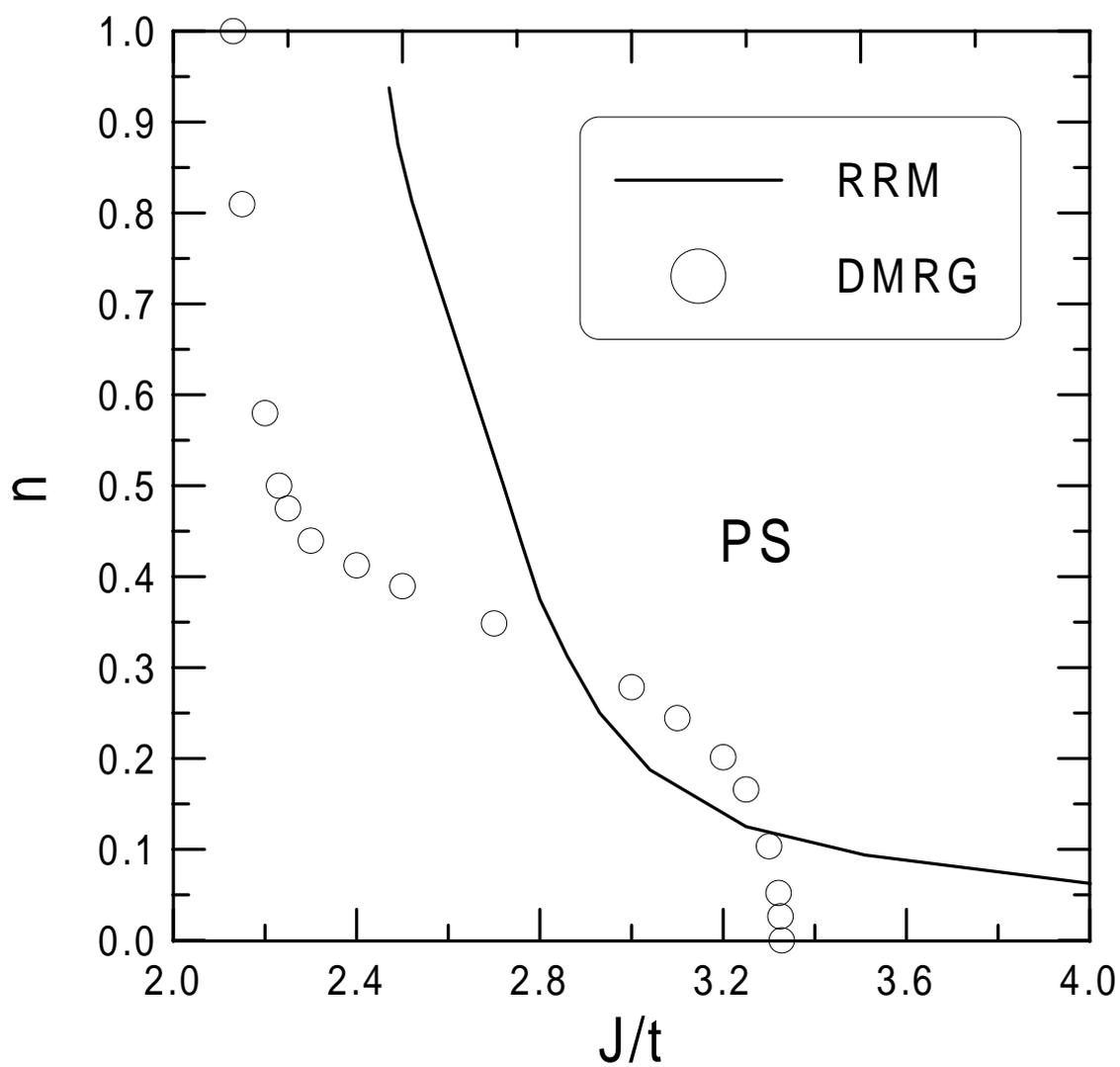
 {\bf Figure 14 :} Boundary of the phase separation region in the case
where $J=J', t=t'$, computed with DMRG and the RRM.

\end{document}